# Automatic Refinement of Force Fields Based on Phase Diagrams

Bin Jin[1], Bin Han[2], Wei Feng[2], Kuang Yu[3*], and Shenzhen Xu[1*]

[1]School of Materials Science and Engineering, Peking University Beijing 100871, China

[2]Tsinghua Shenzhen International Graduate School, Shenzhen 518055, China

[3]ByteDance Seed - AI for Science, Shenzhen Bay Technology Innovation Center, Shenzhen 518000, Guangdong, China

*Corresponding author: xushenzhen@pku.edu.cn, kuangyu.2025@bytedance.com

**Abstract**

Exact characterization of phase transitions requires sufficient configurational sampling, necessitating efficient and accurate potential energy surfaces. Molecular force fields with computational efficiency and physical interpretability are desirable but challenging to refine for complex interactions. To address this, we propose a force field refinement strategy with phase diagrams as top-down optimization targets based on automatic differentiation. Using gas-liquid co-existence as a paradigm, we employ an enhanced sampling technique and design a differentiable loss function to evaluate force fields' depiction of phase diagrams. The refined force fields produce gas-liquid phase diagrams matching well with targets for two modeling systems, which confirms our approach as an effective automated force field development framework for phase transition studies.

Phase transitions play a fundamental role in the realms of thermodynamics and dynamics [1-5]. They occur in regions exhibiting intense fluctuations within a system and represent intrinsic properties of materials. Study and application of phase transitions underpin critical advancements in materials science, energy storage technologies, and condensed matter physics [6-12]. Computational simulations provide a powerful tool for investigating phase transitions [13-17]. However, sufficient sampling of the configurational space is required to accurately describe the process involving different phases. Let's take the simulation of gas-liquid phase co-existence as an example, we may encounter a few challenges regarding the critical point. When the temperature is far from the critical point, fluctuations of physical quantities are relatively small, while the transformation between different phases during the configurational sampling is infrequent, thus requiring a large number of steps to achieve sufficient sampling. Conversely, close to the critical point, the transformation between different phases is frequent in the sampling process, but fluctuations of physical quantities become relatively large. The parallel-tempering Monte Carlo (PTMC) method combines the advantages of sampling under these different thermodynamic conditions to resolve the above challenges [18-20]. Regardless of the sampling methods, an accurate and efficient potential energy surface (PES) is the foundation. There are three common types of PES surrogate models: the classical molecular force field, the machine learning force field (MLFF), and the *ab initio* calculation method. While *ab initio* calculations are accurate, they are computationally expensive for large-scale simulations. Training an MLFF capable of accurately describing the configurational space of interest also presents challenges, which requires sufficient representative samples included in the training dataset to prevent abrupt failures during application of MLFF, and the computational cost inherently scales with the dataset size.

By contrast, the advantages of classical molecular force fields include a high computational efficiency and an inherent physical interpretability of the terms involved in mathematical expressions. However, more complicated expressions with more parameters need to be incorporated in the classical molecular force field to depict complex interactions in various materials' systems more accurately, and the transferability between different systems thus becomes questionable. Too many parameters also lead to difficulty in refining a force field model. Regarding this issue, the automatic differentiation technique has enabled efficient force field parameter refinement, underpinning several packages built based on different platforms, including TorchMD [21], JAX-MD [22], SPONGE [23], DiffTRe [24], and the Differentiable Molecular Force Field (DMFF) [25]. Different from end-to-end differentiable MD engines like TorchMD and JAX-MD, the DMFF platform avoids the gradient vanishing/exploding issues and excessive memory demands in end-to-end differentiation through the entire MD trajectory. Meanwhile, DMFF is applicable for various force field function forms and multiple types of object functions for different physical properties (density, evaporation enthalpy, gas uptake, transport coefficients, spectroscopy and so on) [26-28]. In addition to the classical molecular force field, it can also be extended to the refinement of MLFF, which intrinsically features a large number of parameters [27,29].

The DMFF platform supports the force field development through both bottom-up and top-down approaches. In the top-down strategy, the experimental macroscopic physical quantities of interest are represented by the differentiable ensemble averages from simulations. This strategy can be applied to refine a classical force field for the research on materials' phase transitions. A reliable force field, which could be employed to investigate materials' phase transitions, should at least be able to accurately predict materials' thermodynamic phase diagrams (usually benchmarked with experimental measurements), since the successful construction of a thermodynamic phase diagram is a prerequisite for subsequent research on phase transitions. Therefore, we aim to develop a systematic approach to refine molecular force fields automatically using the DMFF platform, based on pre-acquired phase diagrams. Different thermodynamic conditions from the phase diagrams can be simulated with multiple trajectories simultaneously using the PTMC method. The DMFF platform performs analysis of multiple simulation trajectories under different thermodynamic conditions using the Multistate Bennett Acceptance Ratio (MBAR) algorithm [30]. We highlight that the MBAR reweighting technique is also perfectly suitable for our employed PTMC sampling method with multiple-trajectory simulations.

In this work, we demonstrate the validity and practicality of our proposed approach for automatic refinement of force fields with the gas-liquid phase diagrams as the optimization object. The equilibrium density-temperature ($\rho - T$) phase diagram (**Fig. 1**) is a commonly used form, which exhibits the phases' co-existence behavior.

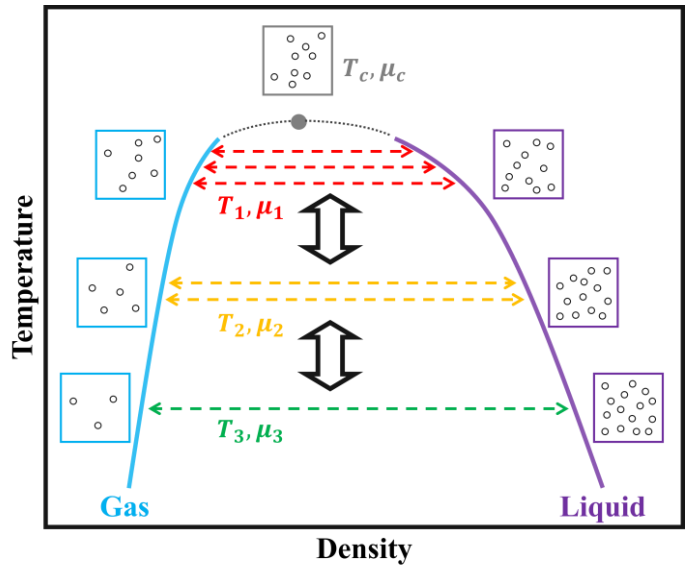

**Fig. 1** Scheme of the gas-liquid phase diagram and the hyper-parallel tempering Monte Carlo (HPTMC) method. The blue and violet curves denote the gas and liquid phases, respectively. The grey solid point denotes the critical point. Boxes with particles illustrate the structures of phases across different

temperatures and associated chemical potentials. The green, yellow and red arrows denote the co-existence of the gas and liquid phases in different replicas. More arrows indicate more frequent transformation between gas and liquid phases during statistical sampling. The hollow black arrows denote the replica exchange.

As for the gas-liquid phase equilibration, the grand canonical ensemble (GCE) is more suitable than the isothermal-isobaric $NPT$ ensemble, since the volume fluctuation would be too large in the sampling of gas-liquid phases' co-existence. We can simulate the GCE by the grand canonical Monte Carlo (GCMC) method [31-33]. Combining the PTMC approach with the GCMC sampling, the hyper-parallel tempering Monte Carlo (HPTMC) method [34,35] was proposed in earlier work to improve the sampling efficiency of the GCE (**Fig. 1**). The basic principle of the HPTMC method is introduced briefly as follows. For a particle-number variable system, the GCE probability density of a state:

$$p(\boldsymbol{R}_N) \propto \frac{1}{\lambda^{3N}} \exp\{-\beta[U(\boldsymbol{R}_N) - \mu N]\} \quad (1)$$

where $\lambda$ is the de Broglie thermal wavelength, $\beta = 1/k_\mathrm{B}T$ is the inverse of temperature, $\boldsymbol{R}_N$ represents the spatial coordinates of the system with $N$ particles, $U(\boldsymbol{R}_N)$ is the system's potential energy, and $\mu$ is the particle's chemical potential. Consider a composite ensemble consisting of $M$ independent replicas of GCEs with different temperatures $\beta_i$ and chemical potentials $\mu_i$. The probability density of a composite state $\{\boldsymbol{R}_{N_i}\}_{i=1}^M$ is:

$$p\left(\{\boldsymbol{R}_{N_i}\}_{i=1}^M\right) \propto \prod_{i=1}^M \frac{1}{\lambda_i^{3N_i}} \exp\{-\beta_i[U(\boldsymbol{R}_{N_i}) - \mu_i N_i]\} \quad (2)$$

The marginal probability density of each replica is the same as its GCE probability density. Two types of trial moves can be used to sample the composite ensemble using the Markov chain:
(1) Update each replica using the GCMC method.
(2) Exchange random pairs of replicas with the index $i$ and $j$, and the acceptance probability is

$$A\left(\boldsymbol{R}_{N_i} \leftrightarrow \boldsymbol{R}_{N_j}\right) = \min\left\{1, \left(\frac{\lambda_j}{\lambda_i}\right)^{3N_j - 3N_i} \exp\left[(\beta_j - \beta_i)\left(U\left(\boldsymbol{R}_{N_j}\right) - U(\boldsymbol{R}_{N_i})\right) - (\beta_j\mu_j - \beta_i\mu_i)(N_j - N_i)\right]\right\}$$

(3)

The HPTMC method integrates these two types of trial moves with a predetermined probability ratio.

Leveraging the thermodynamic ensembles involved in the phase diagrams sampled via the HPTMC method, we establish the workflow of the force field refinement as shown in **Fig. 2**. Initial chemical potentials, force field parameters and the parameters used in our designed loss function (explained by Eq. 6) are set up at the beginning of the refinement. The HPTMC simulation is performed to produce the sampling trajectories. The MBAR method extracts all information from multiple trajectories to produce the density distribution under a target thermodynamic condition. The approximate partition function of each trajectory will be optimized self-consistently (*MBAR optimizing*) each time we do the HPTMC resampling. The weights of all samples under the target thermodynamic condition are then computed (*MBAR reweighting*), and the *reweighted density distribution* is obtained, which is employed to represent **the density profile under this condition**. Subsequently, we can calculate the loss function, and the gradients of which with respect to all parameters are used to update their values. If the criterion of resampling is satisfied, then the above loop (the green-colored loop in **Fig. 2**) is performed again; otherwise, we can simply use the current approximate partition functions to calculate the weights of all samples and continue to update all parameters. (refer to **Supplemental Materials (SM) Section S1** for the details)

The cornerstone of the refinement process lies in the construction of a loss function. The basic idea of our loss function design is: (1) defining an equilibrium density distribution as the reference, which involves information of the target gas and liquid densities; (2) maximizing the similarity between the *reference density distribution* and the *reweighted density distribution* (obtained via the MBAR approach) by adjusting the parameters included in the loss function.

Given $M$ independent trajectories under different ensemble conditions (sampling ensembles) from the HPTMC simulation, each comprising $N_r$ samples, the *reweighted density distribution* of the target ensemble $k$ characterized by a chemical potential $\mu_k$ and an inverse temperature $\beta_k$ is expressed as

$$p_{\text{rwt}}^k(\rho) = \sum_{r=1}^{M} \sum_{j=1}^{N_r} W_{r,j}(\mu_k, \beta_k, \{\theta_i\}) \delta(\rho - \rho_j^r) \qquad (4)$$

where $\rho_j^r$ is the particle density of the sample with the index $j$ within the trajectory $r$, $\{\theta_i\}$ represents the set of all force field parameters, $W_{r,j}(\mu_k, \beta_k, \{\theta_i\})$ denotes the weight of the corresponding sample for the target ensemble $k$. (refer to **SM Subsection S4.3** for the approximation of the Dirac-δ function) Note that the chemical potentials $\{\mu_k\}$ of the target ensembles are also undergoing an optimization process in our refinement framework. We also have $M$ target ensembles, and they share the same temperatures with the corresponding sampling ensembles. The refined chemical potentials $\{\mu_k\}$ will then be used for computing the sampling ensembles in the next loop of HPTMC resampling.

The *reference density distribution* is defined as a bimodal Gaussian distribution, representing the two density peaks associated with the gas phase and the liquid phase, respectively. The areas under two Gaussian peaks are set equal, which approximates a phase equilibrium state.

$$p_{\text{ref}}^k(\rho) = \frac{1}{2\sqrt{2\pi}\sigma_1^k}\exp\left[-\frac{(\rho-\rho_k^g)^2}{2(\sigma_1^k)^2}\right] + \frac{1}{2\sqrt{2\pi}\sigma_2^k}\exp\left[-\frac{(\rho-\rho_k^l)^2}{2(\sigma_2^k)^2}\right] \tag{5}$$

where $\rho_k^g, \rho_k^l$ represent the pre-acquired gas and liquid densities of the target ensemble $k$. The widths of the two Gaussian peaks $\{\sigma_1^k\}, \{\sigma_2^k\}$ will also be automatically refined in our framework.

The similarity between the *reweighted density distribution* and the *reference density distribution* is measured by the Kullback-Leibler divergence [36], then the loss function can be defined as the summation of these similarities at different temperatures:

$$\mathcal{L} = \sum_{k=1}^{M} D_{\text{KL}}\left(p_{\text{rwt}}^k \middle\| p_{\text{ref}}^k\right)$$

$$= \sum_{k=1}^{M} \sum_{\rho \in \varrho} p_{\text{rwt}}^k(\rho) \ln \frac{p_{\text{rwt}}^k(\rho)}{p_{\text{ref}}^k(\rho)}$$

$$= \sum_{k=1}^{M} \sum_{r=1}^{M} \sum_{j=1}^{N_r} W_{r,j}(\mu_k, \beta_k, \{\theta_i\})\left[\ln p_{\text{rwt}}^k(\rho_j^r) - \ln p_{\text{ref}}^k(\rho_j^r)\right] \tag{6}$$

where $D_{\text{KL}}(p\|q)$ is the Kullback-Leibler divergence between two different distributions $p(x)$ and $q(x)$, and $\varrho$ represents the domain of the particle density $\rho$. Since the weights are obtained by the MBAR algorithm, and due to the analytical form of the *reference density distribution*, this loss function is differentiable with respect to the refining parameters. If necessary, a penalty term restraining the widths' difference between the density distribution peaks (representing the two distinct phases) can be added into the loss function. Refer to **SM Subsection S3.3** for the details of the penalty term.

The preceding introduction outlines our phase-diagram-based force field refinement methodology, with novelty residing in three key aspects:
(1) Using the HPTMC method to enhance the sampling efficiency of the phases' equilibration, which is perfectly suitable for the multi-ensemble MBAR reweighting technique employed in our work.
(2) Using the automatic differentiation technique (commonly employed in the artificial intelligence nowadays) to automatically refine the classical force field.

(3) Designing a differentiable loss function to evaluate the capability of a force field for reproducing a gas-liquid phase diagram.

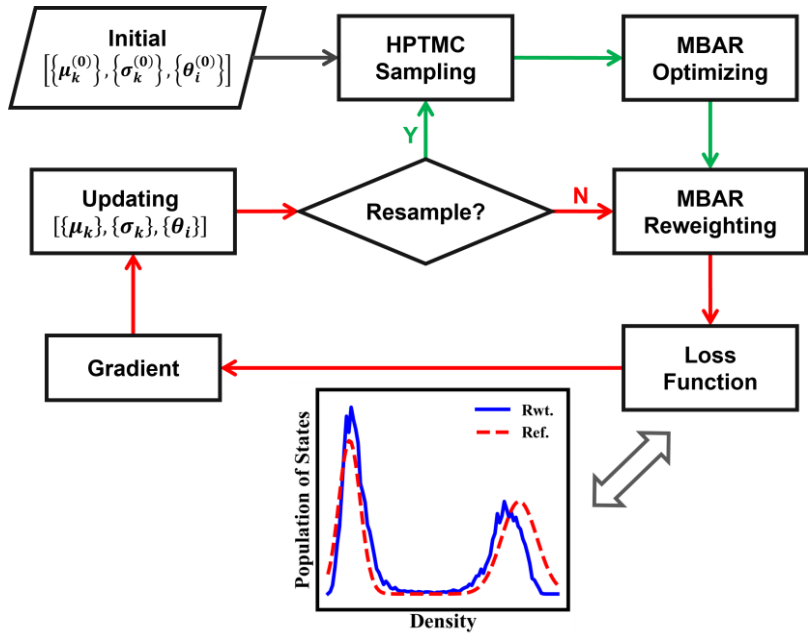

**Fig. 2** Workflow of the force field refinement based on a target phase diagram using the DMFF platform. The red loop reiterates until the resampling of trajectories is necessary. The green loop corresponds to the resampling process. The inset figure illustrates the key idea of the loss function, in which a bimodal Gaussian distribution regarded as the *reference density distribution* is used to compare with the *reweighted density distribution* computed from the sampled trajectories. The similarity of the two distributions is embedded in the loss function.

Two systems, a Lennard-Jones (L-J) potential [37-39] system, and a carbon dioxide ($CO_2$) system (**Fig. 3(a)**) are chosen to demonstrate the validity and efficiency of our proposed phase-diagram-based force field refinement strategy. The L-J potential is a classical model for intramolecular interactions, extensively employed in the simulation across gaseous, liquid and solid materials. The $CO_2$ system has more complicated interactions, with important applications around its critical point in target component extraction, polymer processing, refrigeration, supercritical drying and cleaning and so on [40]. In the L-J potential system, all physical quantities are calculated with the dimensionless arbitrary units (a.u.) for simplicity. A periodic supercell with size $7\times7\times7$ a.u.$^3$ is employed with a potential truncation at 2.5 a.u.. Temperatures used for different replicas of the HPTMC simulations and the *reweighted density distribution* calculations are selected as 1.00, 1.05, 1.10, 1.13, 1.16 a.u.. (refer to **SM Subsection S2.1** for the details of the force field setup and the sampling parameters used in HPTMC simulations) For the $CO_2$ system, a periodic cell with size $20\times20\times20$ Ang$^3$ is considered by using the TraPPE force field [41]. (refer to **SM Subsection S2.2** for the TraPPE force field details and the sampling parameters used in HPTMC simulations) Temperatures for different replicas of the HPTMC simulations and the *reweighted density distribution* calculations are set as 270, 275, 280, 285, 290, 295, 300 K.

To demonstrate our proposed idea of force field refinement, we first pre-select a set of force field parameters for both of the L-J potential system and the $CO_2$ system, and use the HPTMC method to generate the target phase diagrams (the sampled density distributions are shown in **Fig. S2(a, b)**, the sampling efficiency validated by the exchange frequency among different replicas is presented in **Fig. S2(c, d)**). We then apply perturbations to the force field parameters, and employ our developed refinement framework to see if the parameters can be optimized toward the original values. The number of samples, optimization methods, learning rates, optimization steps and resampling interval used in the refinement process of the L-J potential system and the $CO_2$ system can be referred to **SM Section S3**. We note that the preset force field parameters of the L-J potential system and the $CO_2$ system are obtained from previous literatures [34,41]. For the $CO_2$ system, in addition to the target $\rho - T$ diagram generated by our simulations, we can also use the experimental $CO_2$ gas-liquid phase diagram [42] as a target to justify the effectiveness of our developed force field refinement approach.

We can see in **Fig. 3(b - d)** that the initially predicted gas and liquid densities by the perturbed force field before refinement deviate significantly from the target values across our studied temperature range. We also note that, these initial density points' trend with respect to temperatures looks quite unsmooth, primarily due to two reasons: (1) the chemical potentials employed in the initial simulations are not yet optimized to achieve exact phase equilibrium, causing errors for the positions of the density profile's peaks compared to the truly equilibrium state; (2) phase densities are estimated by fitting the *reweighted density distribution* with a bimodal Gaussian distribution (refer to **SM Subsection S4.4** for details of the fitting process). Consequently, when the system is relatively far from a phase equilibrium state, the density distribution may not exhibit an obvious two-peak feature, which could lead to bias when fitted with a bimodal Gaussian distribution. As the force field refinement calculations proceed, the chemical potentials will approach the values for phase equilibrium, and the two-peak features of the density profiles also become more obvious, leading to better fitting with the bimodal Gaussian distribution. This trend can be reflected by the evolvement of *reweighted density distributions* along the force field refinement process as shown in **SM Fig. S3 - S5**.

After the force field refinement gets converged, the co-existence curves computed with the refined force field match well with the targets as shown in **Fig. 3(b - d)**. For the $CO_2$ system with the experimental phase diagram as the target, our refinement procedure is indeed effective (as shown in **Fig. 3(d),** and **Fig. S5** for the evolvement of *reweighted density distribution* during the refinement), although not achieving a quantitatively perfect match. We need to point out that the representational capacity of the TraPPE force field is constrained by its limited number of parameters, and the phase densities estimated by the bimodal Gaussian distribution may also exhibit certain numerical deviations from the true values. We also see in **Fig. 3(d)** that the critical point predicted with the refined force field is close to the experimental one. We need to clarify that the information of the experimental critical point is not included in our force field refinement calculations, this comparison thus indicates a certain transferability of the force field refined by our proposed workflow in terms of predicting the system's thermodynamic properties. We note here that the computed critical-point temperature might be overestimated due to the strict criteria of identifying a continuous phase transition (refer to **SM Subsection S4.2** for details) introduced by us to set up a quantitative standard. Overall, the results demonstrate that our developed force field refinement strategy is applicable to multiple material systems with different forms of force fields, including a toy model of L-J potential and a more practical modeling system of $CO_2$ across the gas and liquid phases.

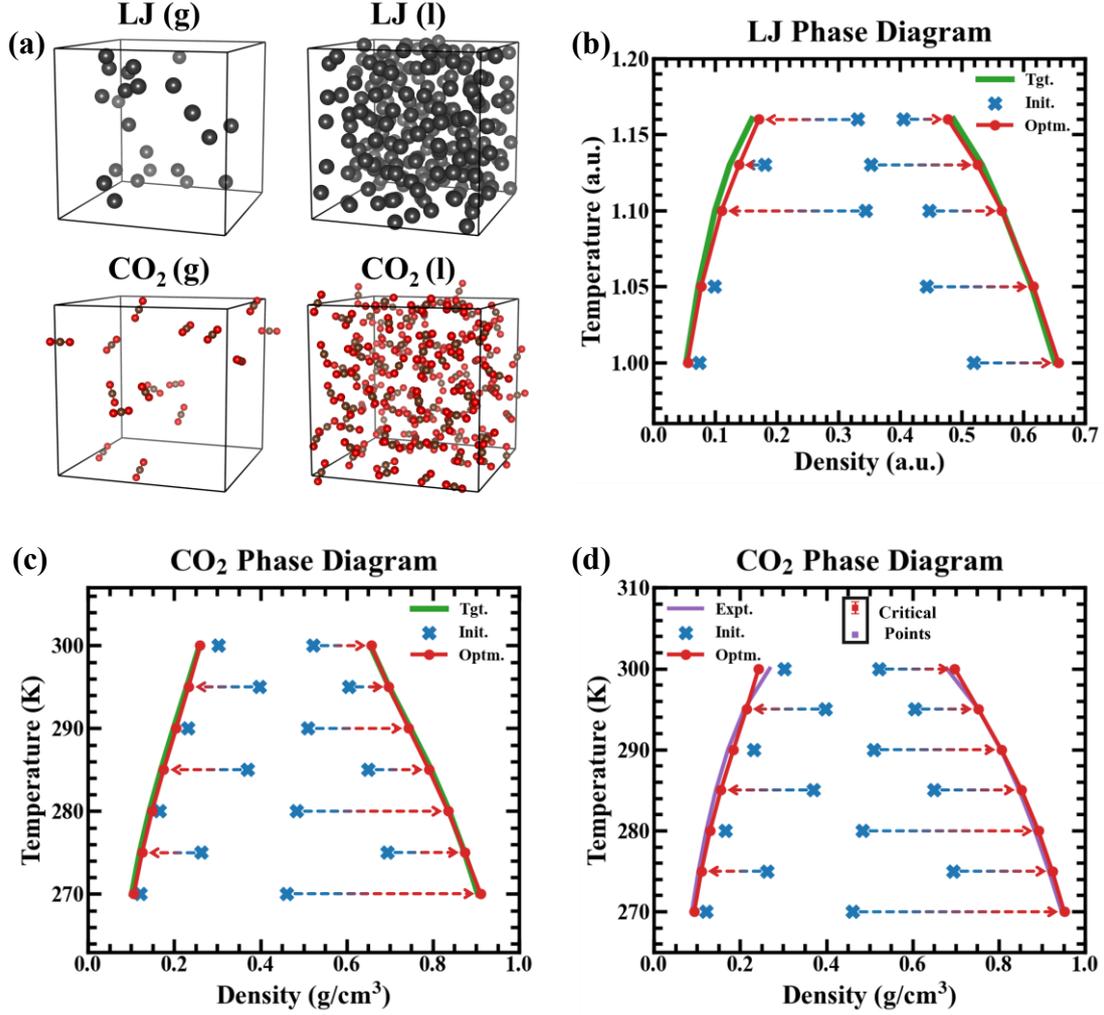

**Fig. 3** Structures of gas and liquid phases for the L-J potential system and the $CO_2$ system (a), and the results of phase diagram optimizations with the force field refinement using the DMFF platform for the L-J potential system (b) and the $CO_2$ system (c)(d). The blue crosses and the red lines with circle points correspond to the computed phase co-existence curves using the HPTMC method with the initial force field and the refined force field, respectively. The green and violet lines denote the targeted co-existence curves, the green lines correspond to simulation results with preset force field parameters, and the violet lines are from experiments. The dashed arrows illustrate the direction of the density variation during the refinement process. The critical points of the $CO_2$ system from the experiment and our simulations using the refined force field are shown in (d).

In addition to the above results, several methodological insights merit further discussion. During the refinement process, the initially perturbed force field parameters are refined towards the target values effectively, and the sensitivity of the phase co-existence $\rho - T$ curve on each parameter can be reflected by the parameters' variation (shown in **Fig. S6**). As for the $CO_2$ system, refinement of the parameter $\sigma_O$ is the most effective, indicating that the performance of the TraPPE force field on the phase diagram prediction sensitively depends on the parameter $\sigma_O$, which possibly stems from the fact that oxygen atoms are located at the two ends of the linear $CO_2$ molecule. Such an example shows that analyzing the sensitivity of force field parameters offers valuable insights into the underlying interactions of the material system. While our strategy is primarily designed for force field parameters, it concurrently refines the chemical potentials. This approach enables efficient and accurate determination of phase-equilibrium chemical potentials, which is otherwise unfeasible by manual adjustment. We further realize that the density distribution from the MBAR reweighted sampling scheme typically has peaks with a slightly asymmetric feature under the phase co-existence condition, so this asymmetry may not be well captured by the symmetric Gaussian distribution. Skewed Gaussian distribution could be employed to address this limitation in the future. Another potential drawback is that multiple force field parameters in one type of potential energy function are sometimes interrelated when predicting a single property (e.g. the gas-liquid phase diagram in this work), which results in difficulty of independent refinement for these multiple parameters. This challenge could be resolved through refinement with multi-objects (i.e. physical properties) as targets. We will try to extend our developed strategy to automatic refinement of force fields based on solid-liquid phase diagrams in the future. This approach can also be implemented for the refinement of MLFFs. For example, an initial MLFF first can be obtained by training with the *ab-initio* dataset, we then can fine tune the parameters of this MLFF by fitting the investigated material system's phase diagrams via the top-down framework in this work.

In summary, we develop a phase-diagram-based force field refinement method leveraging automatic differentiation technique via the DMFF platform, to generate reliable classical force fields for characterizing phases' co-existence. HPTMC methods are applied to realize sufficient phases' equilibration sampling. The strategy is validated on the gas-liquid phase diagram, for which a differentiable loss function is designed. The approach yields refined force fields producing phase diagrams closely matching with targets, while automatically determining phase-equilibrium chemical potentials. Our proposed methodology thus establishes an extensible paradigm for refining force fields dealing with phase transitions.

The authors acknowledge funding support from the National Natural Science Foundation of China (grant no. 92470114, no. 52273223), Ministry of Science and Technology of the People's Republic of China (grant no. 2021YFB3800303), and DP Technology Corporation (grant no. 2021110016001141). The computing resource of this work was provided by the Bohrium Cloud Platform, which was supported by DP Technology.

# Supplemental Material for
# Automatic Refinement of Force Fields Based on Phase Diagrams

Bin Jin[1], Bin Han[2], Wei Feng[2], Kuang Yu[3*], and Shenzhen Xu[1*]

[1]School of Materials Science and Engineering, Peking University Beijing 100871, China

[2]Tsinghua Shenzhen International Graduate School, Shenzhen 518055, China

[3]ByteDance Seed - AI for Science, Shenzhen Bay Technology Innovation Center, Shenzhen 518000, Guangdong, China

*Corresponding author: xushenzhen@pku.edu.cn, kuangyu.2025@bytedance.com

## S1. Details of the Multistate Bennett Acceptance Ratio Method

Forms of the effective energy $u(\boldsymbol{R})$ of a system (particles' coordinates as $\boldsymbol{R}$) are different in different ensembles. If $U(\boldsymbol{R})$ is the system's potential energy, and $P, V(\boldsymbol{R}), \mu, N(\boldsymbol{R})$ are the pressure, volume, chemical potential and particle number of the system, then

$$u(\boldsymbol{R}) = \begin{cases} U(\boldsymbol{R}), & \text{for } NVT \text{ ensemble} \\ U(\boldsymbol{R}) + PV(\boldsymbol{R}), & \text{for } NPT \text{ ensemble} \\ U(\boldsymbol{R}) - \mu N(\boldsymbol{R}), & \text{for } \mu VT \text{ ensemble} \end{cases}$$

(S1)

In the MBAR method, considering $K$ ensembles with the effective energies $u_k(\boldsymbol{R}), (k = 1, \ldots, K)$. The probability of the state $\boldsymbol{R}$ in the ensemble $k$ is proportional to $w_k(\boldsymbol{R}) = \exp[-\beta_k u_k(\boldsymbol{R})]$, and the corresponding configurational partition function is

$$\mathcal{Z}_k = \int d\boldsymbol{R} w_k(\boldsymbol{R}) \qquad (S2)$$

Assume that the number of samples in the trajectory of the ensemble $k$ is $N_k$, and the sample with the index $n$ within the trajectory is $\boldsymbol{R}_{k,n}$. Then according to the MBAR method, the statistically optimal approximate partition function of these $K$ ensembles satisfy the following self-consistent equations

$$\mathcal{Z}_l = \sum_{k=1}^{K} \sum_{n=1}^{N_k} \frac{w_l(\boldsymbol{R}_{k,n})}{\sum_{r=1}^{K} N_r \mathcal{Z}_r^{-1} w_r(\boldsymbol{R}_{k,n})}, \qquad l = 1, \ldots, K$$

(S3)

Since these equations are nonlinear, relative values of $\{\mathcal{Z}_k\}_{k=1}^{K}$ have to be solved iteratively. This process is called *MBAR optimizing*. These $K$ ensembles can be named as *sampling ensembles*. The other ensembles involved in the statistical computation of our interested physical properties are named as *target ensembles*, which are typically decoupled from the above *sampling ensembles*, and the differentiation operations with respect to force field parameters are applied onto the *target ensemble* averages [1].

For a *target ensemble* with the index $q$, if its probability density is proportional to $w_q(\boldsymbol{R})$, the approximate partition function satisfies the similar set of equations:

$$\mathcal{Z}_q = \sum_{k=1}^{K} \sum_{n=1}^{N_k} \frac{w_q(\boldsymbol{R}_{k,n})}{\sum_{r=1}^{K} N_r \mathcal{Z}_r^{-1} w_r(\boldsymbol{R}_{k,n})} \tag{S4}$$

Here we note that $\mathcal{Z}_q$ is decoupled from $\{\mathcal{Z}_k\}_{k=1}^{K}$, since there is no sample of the ensemble $q$ in the $K$ trajectories. The un-normalized weight of the sample $\boldsymbol{R}_{k,n}$ under the ensemble $q$ can be defined as:

$$p_{k,n}^q = \frac{w_q(\boldsymbol{R}_{k,n})}{\sum_{r=1}^{K} N_r \mathcal{Z}_r^{-1} w_r(\boldsymbol{R}_{k,n})} \tag{S5}$$

and the normalized weight is:

$$W_{k,n}^q = \frac{p_{k,n}^q}{\mathcal{Z}_q} \tag{S6}$$

This process is called *MBAR reweighting*.

Finally, the ensemble average of a physical quantity $A(\boldsymbol{R})$ under the condition of ensemble $q$ can be expressed as:

$$\langle A \rangle_q = \sum_{k=1}^{K} \sum_{n=1}^{N_k} W_{k,n}^q A(\boldsymbol{R}_{k,n})$$

$$\tag{S7}$$

## S2. Sampling Details

### S2.1 L-J Potential System

The truncated L-J potential with $r_{\text{cut}} = 2.5$ a.u. is expressed as [2]

$$V(r) = 4\varepsilon \left[ \left(\frac{\sigma}{r}\right)^{12} - \left(\frac{\sigma}{r}\right)^{6} \right]$$

$$V_{\text{trunc}}(r) = \begin{cases} V(r), & r \leq r_{\text{cut}} \\ 0, & r > r_{\text{cut}} \end{cases}$$

$$\tag{S8}$$

Parameters undergoing refinement in this force field are $\varepsilon, \sigma$.

**Table S1** Key parameters of HPTMC simulations for the L-J potential system.

| Parameter | Time step | GCMC interval | HPTMC interval | Sample interval | Steps | Samples |
|---|---|---|---|---|---|---|
| Value | 0.001 a.u. | 10 | 100 | 1,000 | 2,000,000 | 5×2,000 |

Brief explanations for the parameters:
- Time step: molecular dynamics (MD) time step used to evolve the particles' coordinates under the Nose-Hoover thermostat in this work.
- GCMC interval: number of configurational evolution MD steps between two adjacent particle number variation trial moves during GCMC sampling for each replica.
- HPTMC interval: number of GCMC steps between two adjacent replica exchange trial moves.
- Sample interval: number of simulation steps between two adjacent outputs of samples along our hyper parallel-tempering GCMC trajectories.
- Steps: total number of simulation steps.
- Samples: total number of samples obtained from all replicas, where 5×2,000 means 5 trajectories, each with 2,000 samples.

**S2.2 $CO_2$ System**

The TraPPE force field [3] consists of an L-J potential truncated at 10.0 Ang and a Coulomb potential calculated with the Ewald summation [4] with a cutoff at 10.0 Ang. The truncated L-J potential owns a similar form as Eq. S8. However, C and O have their own parameters $\varepsilon_C, \varepsilon_O, \sigma_C, \sigma_O$, and the parameters of the interaction between C and O are set as

$$\varepsilon_{\text{mix}} = \sqrt{\varepsilon_C \varepsilon_O}$$

$$\sigma_{\text{mix}} = \frac{1}{2}(\sigma_C + \sigma_O)$$

$$\tag{S9}$$

A Coulomb potential

$$V_{\text{coul}}(r) = \frac{C q_i q_j}{\epsilon r}$$

$$\tag{S10}$$

where $C$ is an energy-conversion constant, $q_i, q_j$ are atomic charges, and $\epsilon = 1.0$ is the dielectric constant. A standard Ewald summation is employed to calculate the Coulomb potential. The Coulombic cutoff specified for this potential means that pairwise interactions within this distance are computed directly, and interactions beyond this distance are computed in reciprocal space. Tail correction is attached to add a long-range correction to the energy [2]. Parameters undergoing refinement in this force field are $\varepsilon_C, \varepsilon_O, \sigma_C, \sigma_O$. Charges of the C and O atoms are set to be 0.70 e, −0.35 e, respectively, and they are fixed along the refinement process [3]. $CO_2$ molecules are treated as rigid bodies with the C-O bond length set as 1.16 Ang (default setup) [3].

**Table S2** Key parameters of HPTMC simulations for the $CO_2$ system. Meanings of the parameters are the same as those in the above **Table S1**.

| Parameter | Time step | GCMC interval | HPTMC interval | Sample interval | Steps | Samples |
|---|---|---|---|---|---|---|
| **Value** | 1.0 fs | 100 | 100 | 1,000 | 2,000,000 | 7×2,000 |

## S3. Details of Our Force Field Refinement Workflow

### S3.1 L-J Potential System

**Table S3** Key parameters of force field refinement for the L-J potential system.

| Parameter | Samples | Optimization method | Learning rate | Optimization steps | Resampling interval |
|---|---|---|---|---|---|
| **Value** | 5×800 | Adam [5] | 0.001 | 800 | 50 |

Brief explanations for the parameters:
- Samples: number of samples used in the refinement, and 5×800 means 5 trajectories, each with 800 samples, which are extracted from the sampling trajectories.
- Optimization method: the method employed to update parameters by computing gradients.
- Learning rate: the learning rate of the optimization method.
- Optimization steps: total number of steps for parameters optimization.
- Resampling interval: the number of optimization steps between two adjacent resampling processes.

## S3.2 CO₂ System

**Table S4** Key parameters of force field refinement for the CO$_2$ system. Meanings of the parameters are the same as those in the above **Table S3**.

| Parameter | Samples | Optimization method | Learning rate | Optimization steps | Resampling interval |
|---|---|---|---|---|---|
| **Value** | 7×600 | Adam [5] | 0.001 | 400 for Tgt. / 840 for Expt. | 40 |

## S3.3 Penalty Term Restraining the Widths' Difference between Density Distribution Peaks in the Loss Function

When the system is far from phase equilibrium, the density distribution may exhibit a dominant peak associated with the primary phase alongside a broadened peak related to the other phase. This density profile can mislead the fitting of the bimodal Gaussian distribution in reference density distribution. Specifically, even though the areas under two Gaussian peaks are constrained to be equal, the width of one peak may become artificially enlarged to accommodate the broadened minority peak. This distortion leads the system away from the true phase equilibrium state. To mitigate this issue, a penalty term can be incorporated into the loss function to restrain the peak widths' difference in the bimodal Gaussian expression during refinement.

$$\mathcal{L}_{\text{penalty}} = \sum_{k=1}^{M} \exp\left[200\left(\sigma_1^k - \sigma_2^k\right)^4\right]$$

(S11)

We can see in the above form that, when the widths' difference between the two Gaussian peaks increases, there would be a significant penalty applied onto the total loss function. The parameters in the exponential terms are chosen based on our tests for the performance of the force field refinement calculations.

## S4. Details of Extracting Density Distribution and Relevant Results

### S4.1 Direct Output of Sampled Density Distribution

The direct results of sampled density distribution of a trajectory with $N$ samples are defined as (without the calculation step of the MBAR reweighting):

$$p_{\text{direct}}(\rho) = \sum_{j=1}^{N} \delta(\rho - \rho_j)$$

(S12)

In the numerical calculation, the Dirac-δ function is approximated with a Gaussian distribution. The $\sigma$ of the Gaussian distribution is 0.005 a.u. for the L-J potential system, and 0.010 g/cm$^3$ for the $CO_2$ system. The sampled density distribution is used to determine the critical point of the $CO_2$ system and obtain the phase-equilibrium density profiles of two phases. (refer to **SM Subsection S4.2** and **SM Section S5** for the reasons of using the sampled density distribution in these two situations, respectively)

**S4.2 Simulation of the Critical Point of the $CO_2$ System**

Critical point of the $CO_2$ system represented by the refined force field is obtained by performing HPTMC simulations at different temperatures around the experimental critical point. For each HPTMC simulation, all replicas share the same temperature and have different chemical potentials. We select the replica closest to the phase equilibrium state according to the density distributions, and then determine the temperature of the critical point based on the equilibrium density profile's feature.

We fit the density distribution with a bimodal Gaussian model and there will be three extrema, two local-maximum points corresponding to the two peaks related to gas and liquid phases respectively, and one local-minimum point between the two peaks. Then two criteria are considered to judge if the two phases of gas and liquid cannot be distinguished (i.e., if it is a continuous phase transition):

(1) The probability density of the local-minimum point is larger than half of the lower local-maximum point.
(2) The probability density of the local-minimum point is larger than half of the higher local-maximum point.

If both criteria are met, the two phases at that temperature are considered as indistinguishable. Conversely, if neither criterion is satisfied, we treat the system as in a state with two distinct phases. The temperature range where only one criterion is satisfied thus defines the bounds of uncertainty for our computed critical point.

Sampled density distributions (refer to **SM Subsection S4.1** for the details of the sampled density distribution) of four different temperatures around the experimental critical point are shown in **Fig. S1**. Here we explain that, since the chemical potentials of different replicas are sufficiently close, it allows us to directly use the sampled density distribution, without the need for computing the reweighted density distribution further. According to the above criteria for determining the critical point, the lower and upper bound of the critical point's uncertainty region are selected as $T = 307\,\text{K}$ and $T = 308\,\text{K}$, respectively. And the density corresponding to the critical point is calculated as the average of gas and liquid phase densities at both temperatures.

Since the local-maximum points are overestimated and the local-minimum point is underestimated by the fitted bimodal Gaussian distribution, as we can see in **Fig. S1**, even at $T = 306\,\text{K}$, the sampled density distribution shows significant overlap between the peaks associated with the gas and liquid phases. Our criteria for identifying the continuous phase transition therefore might be too strict, leading to an overestimation of the critical point. Nevertheless, a quantitative standard is still required to determine the critical region's bounds, hence we maintain the fore-mentioned criteria. Adopting alternative "loose" criteria is highly likely to yield a critical point with a lower temperature, thus matching better with the experimental values.

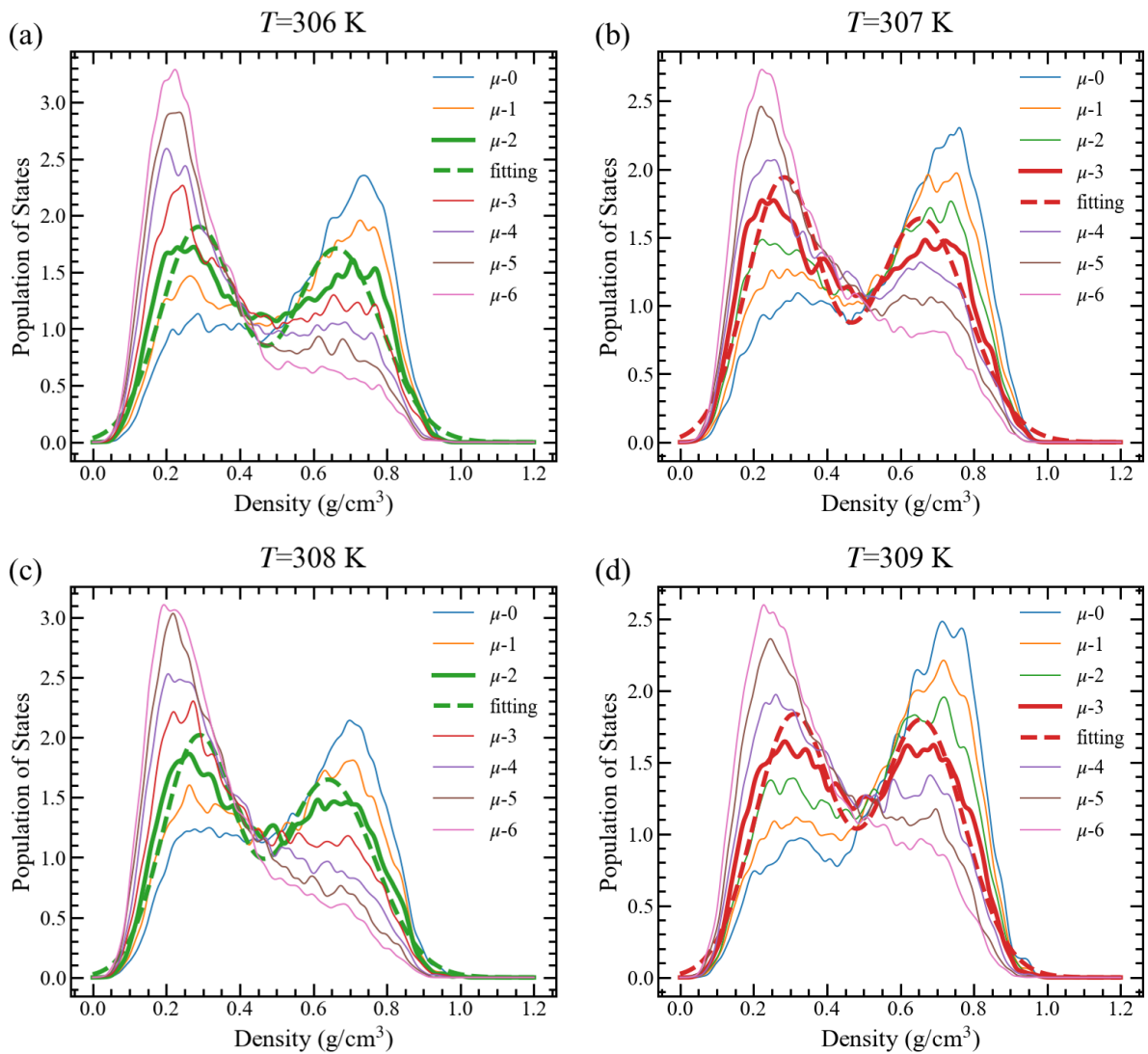

**Fig. S1** Sampled density distributions of the CO$_2$ system simulated with the HPTMC method using the refined force field with the experimental phase diagram as the target. Each panel is related to a HPTMC simulation with all replicas sharing the same temperature, but having different chemical potentials (represented by the different color lines in each panel). The highlighted bold line denotes the chemical potential closest to the phase equilibrium state and the dashed line represents the corresponding density profile fitted with the bimodal Gaussian distribution.

### S4.3 Reweighted Density Distribution

Different from the direct output of sampled density distribution, the reweighted density distribution is calculated using the MBAR method with all trajectories from the HPTMC simulations using Eq. 4 in the main text. In numerical calculations, the Dirac-δ function is approximated with a Gaussian distribution, which is consistent with the approach introduced in **SM Subsection S4.1**.

### S4.4 Estimation of the Phase Density from Reweighted Density Distribution

Similar to the strategy of the loss function's construction, a bimodal Gaussian distribution is used to fit the reweighted density distribution using the expectation-maximization algorithm [6].

$$p_{\text{fitting}}(\rho) = \frac{\pi_1}{\sqrt{2\pi}\sigma_1}\exp\left[-\frac{(\rho-\rho^{\text{g}})^2}{2\sigma_1^2}\right] + \frac{\pi_2}{\sqrt{2\pi}\sigma_2}\exp\left[-\frac{(\rho-\rho^{\text{l}})^2}{2\sigma_2^2}\right]$$

(S13)

A difference from the reference density distribution (Eq. 5 in the main text) employed in our designed loss function is that the areas of the two Gaussian peaks here are not set to be the same. The fitted $\rho^{\text{g}}, \rho^{\text{l}}$ parameters are just the estimated gas and liquid phase densities as the outputs.

## S5. Miscellaneous Results of HPTMC Simulations and DMFF Refinement

Under a phase equilibrium condition, areas of the gas and liquid density peaks are almost the same. As the temperature increases, these two phases become less distinguishable, and the density peaks of them tend to approach each other, as shown in **Fig. S2(a, b)**. We use the sampled density distribution to directly capture the phase-equilibrium density profiles of the target force field, since it is more convenient than the reweighed density distribution and contains enough information to reflect the density variation trend of two phases under different temperatures.

The sampling efficiency of the HPTMC method can be reflected by the replica exchange frequency. For both systems, a representative thermodynamic condition shuttles among different replicas frequently and sufficiently as shown in **Fig. S2(c, d)**, confirming the HPTMC's efficacy in exchanging the configurations across different thermodynamic conditions and enhancing the phase-equilibration sampling.

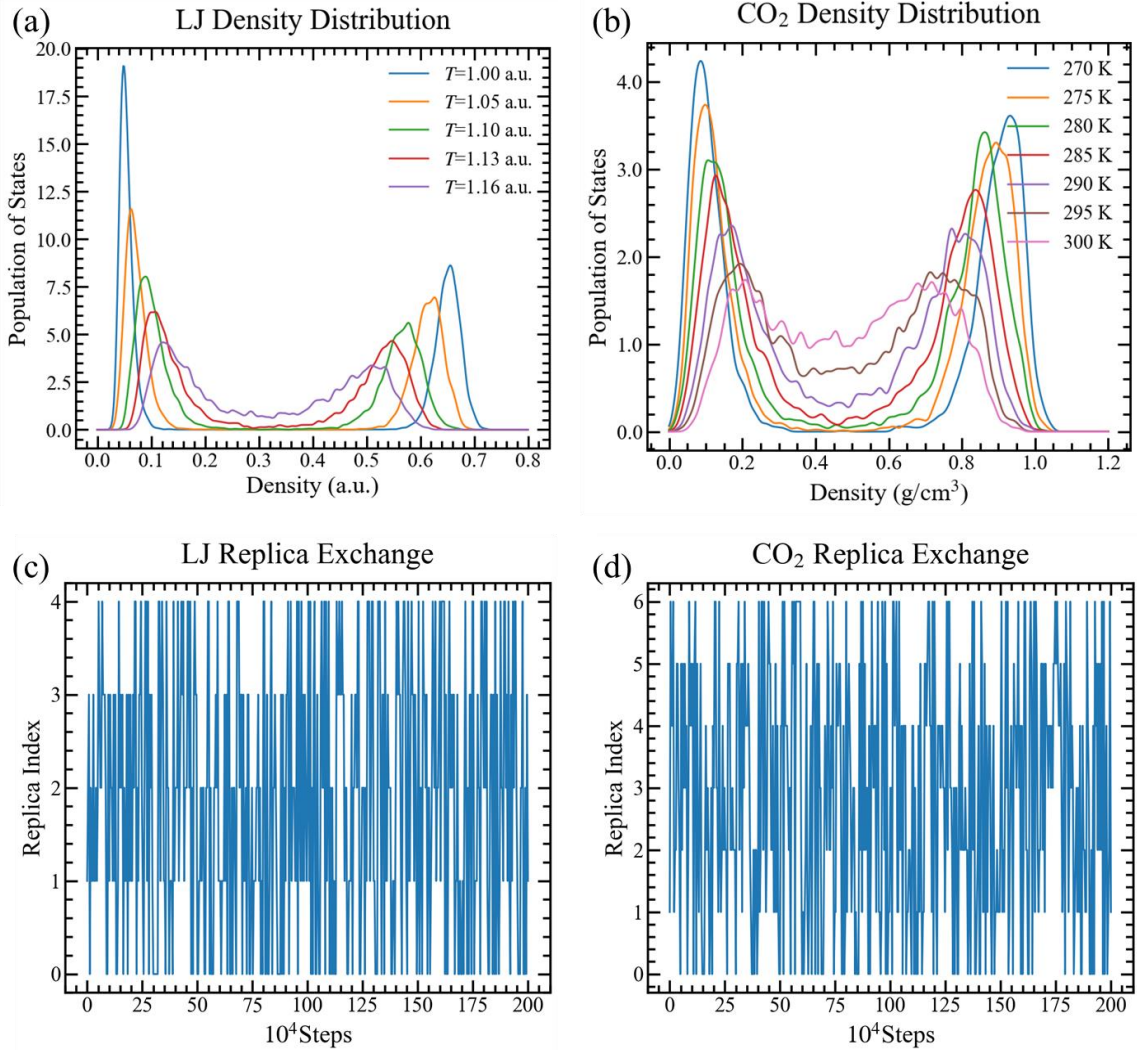

**Fig. S2** Phase equilibrium sampled density distribution of the target force field simulated with the HPTMC method for the L-J potential system (a) and the $CO_2$ system (b), and the variation of replica index with respect to simulation steps for the L-J potential system (c) and the $CO_2$ system (d) under a thermodynamic condition relatively far from the critical point ($T$=1.05 a.u. for the L-J potential system, and 275 K for the $CO_2$ system). Different colored curves in panel (a) and (b) correspond to different thermodynamic conditions described by the temperature and the associated phase-equilibrium chemical potential.

The reweighted density distribution is optimized along the refinement of the force field as shown in **Fig. S3-S5**. For both systems, force field refinement drives the gas and liquid density peak positions to approach the target values, phase density peak areas equalizing and the two-peak feature of the density profile also become more obvious.

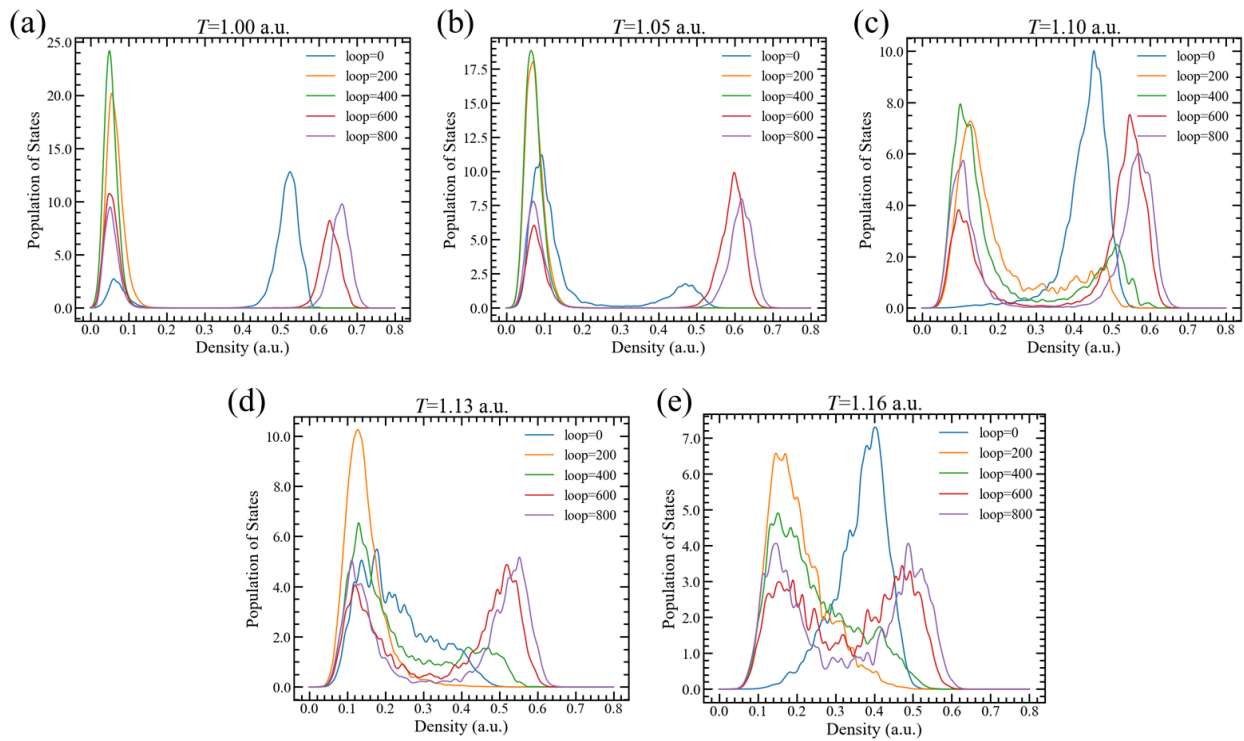

**Fig. S3** Evolvement of reweighted density distributions along the force field refinement process for the L-J potential system under different temperatures (a) 1.00 a.u., (b) 1.05 a.u., (c) 1.10 a.u., (d) 1.13 a.u., (e) 1.16 a.u.. The density distribution is calculated by the MBAR method using the trajectories simulated by the HPTMC method. Different colored lines correspond to the results at different optimization loops.

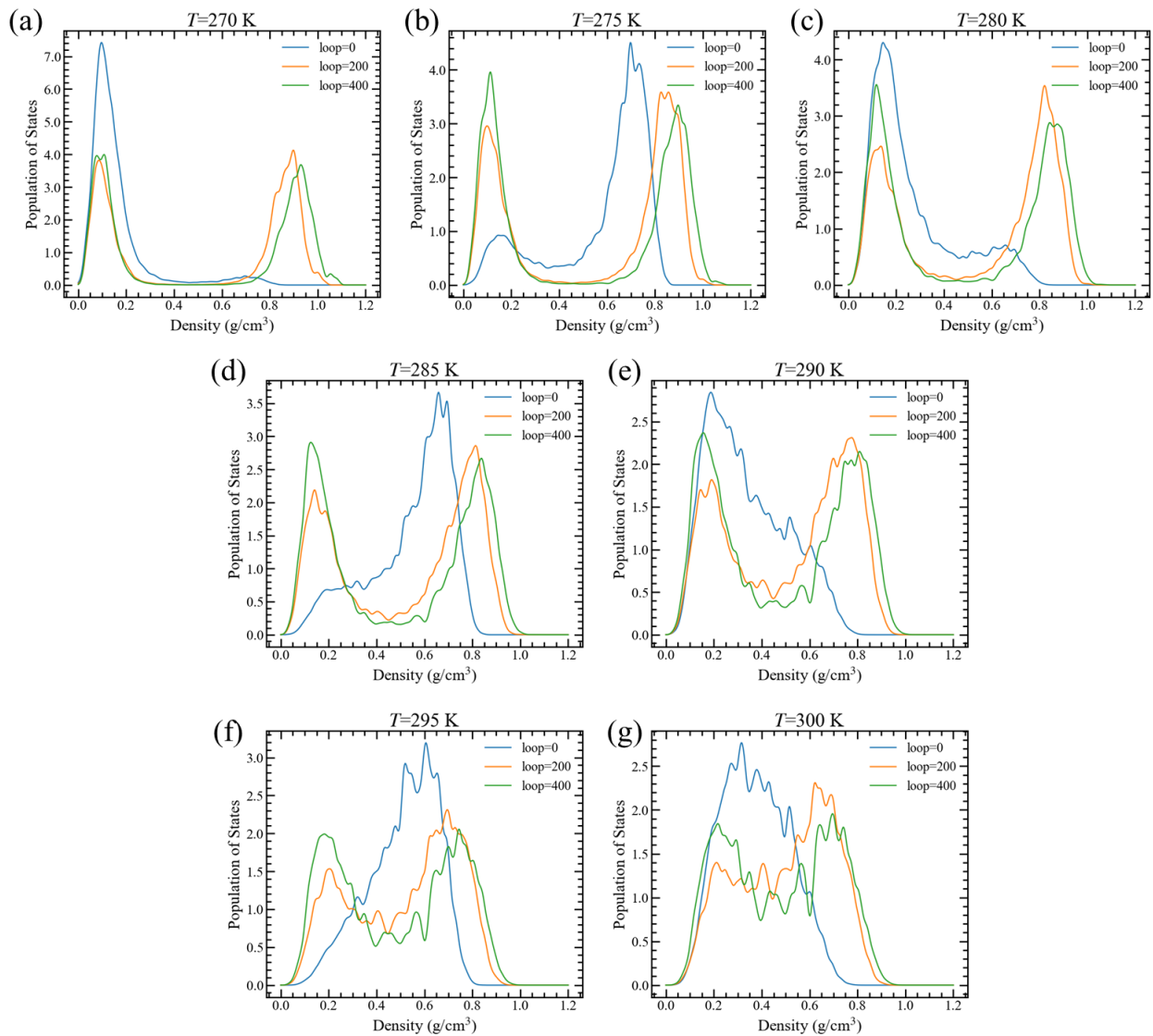

**Fig. S4** Evolvement of reweighted density distributions along the force field refinement process with the simulated phase diagram as the target for the $CO_2$ system under different temperatures (a) 270 K, (b) 275 K, (c) 280 K, (d) 285 K, (e) 290 K, (f) 295 K, (g) 300 K. The density distribution is calculated by the MBAR method using the trajectories simulated by the HPTMC method. Different colored lines correspond to the results at different optimization loops.

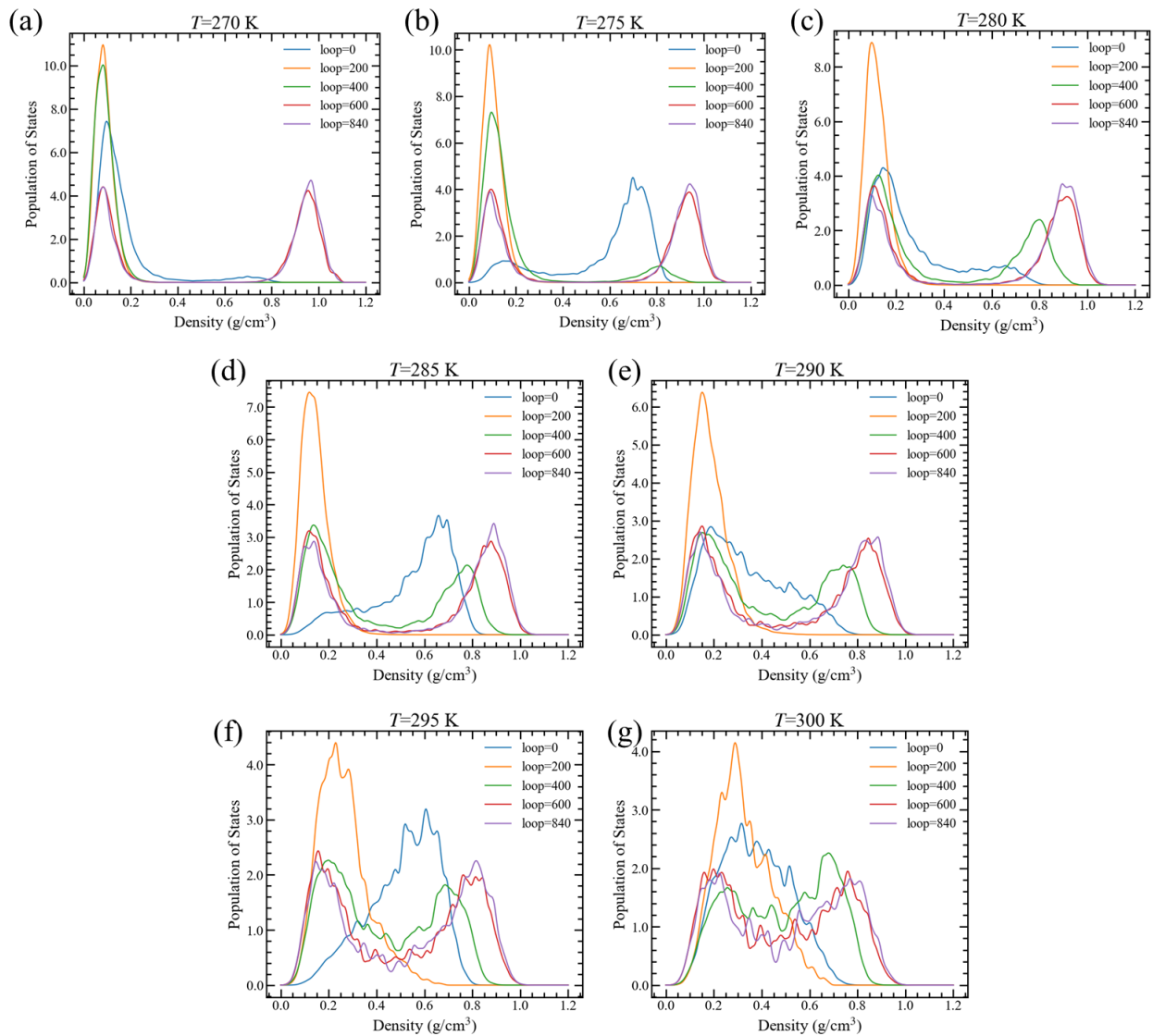

**Fig. S5** Evolvement of reweighted density distributions along the force field refinement process with the experimental phase diagram as the target for the $CO_2$ system under different temperatures (a) 270 K, (b) 275 K, (c) 280 K, (d) 285 K, (e) 290 K, (f) 295 K, (g) 300 K. The density distribution is calculated by the MBAR method using the trajectories simulated by the HPTMC method. Different colored lines correspond to the results at different optimization loops.

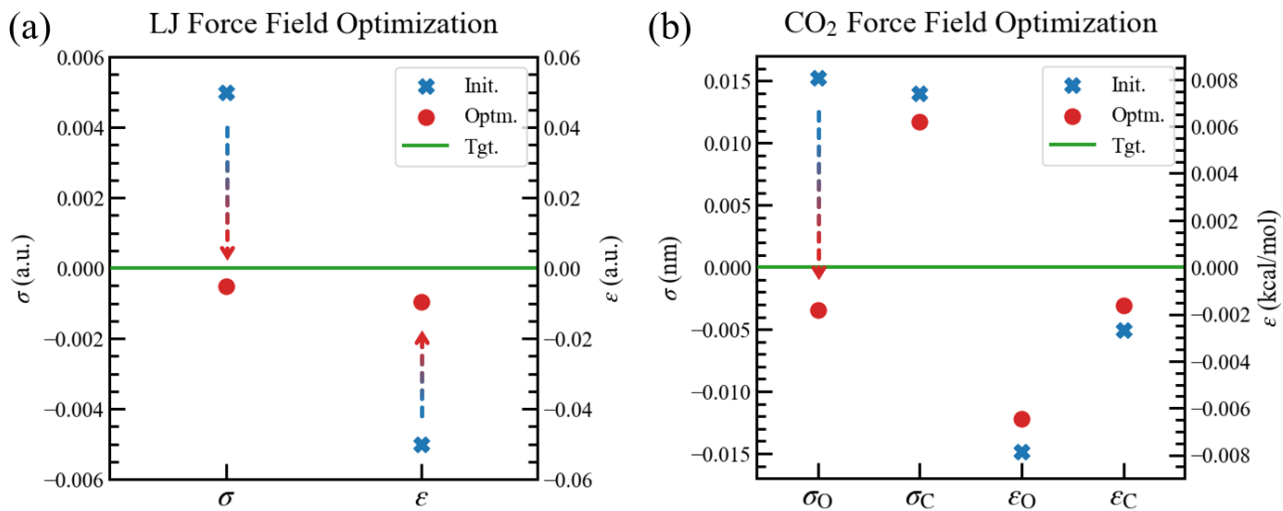

**Fig. S6** Force field parameters' variations along the refinement process for the L-J potential system (a) and the $CO_2$ system (b). The green line denotes the target values for the preset force field parameters (set as zero for the simplicity of comparison). The blue crosses and the red circles correspond to the initial force field parameters (after perturbation) and the refined force field parameters with respect to the target ones, respectively. The dashed arrows illustrate the direction of the parameter variation during the refinement.